\documentclass{article}
\usepackage{lipsum}
\usepackage{authblk}
\usepackage[english]{babel}
\usepackage{amsfonts}
\usepackage{amsmath}
\usepackage[top=2cm, bottom=2cm, left=2cm, right=2cm]{geometry}
\usepackage{fancyhdr}
\usepackage{multicol}
\usepackage{epsfig}
\usepackage{graphics}
\usepackage{epsfig}
\usepackage{lipsum}
\usepackage{fancyhdr}

\renewenvironment{abstract}{
	
	\hfill\begin{minipage}{0.95\textwidth}
		\rule{\textwidth}{1pt}}
		{\par\noindent\rule{\textwidth}{1pt}\end{minipage}
	}

\newcommand{\ket}[1]{\bigl| #1 \bigr\rangle}


\begin{document}
%
\title{Teleportation two-qubit state by using  two different  protocols }
\author[1]{ \textbf{K. El anouz}}
\author[1,2]{ \textbf{A. El Allati}}
\author[3,4]{ \textbf{N. Metwally}}
\affil[1]{\small Laboratory of R\&D in Engineering Sciences, Faculty of Sciences and Techniques Al-Hoceima, Abdelmalek Essaadi University, Tetouan,
	Morocco}
\affil[2]{\small  The Abdus Salam International Center for Theoretical Physics,
	Strada Costiera 11, Miramare-Trieste, Italy} \affil[3]{\small Math. Dept.,
	College of Science, University of Bahrain,  P. O. Box 32038
	Kingdom of Bahrain} \affil[4]{\small Math. Dept. Faculty of
	Science, Aswan University, Aswan, Egypt}
\maketitle
\begin{abstract}
In this contribution, two versions of teleportation protocol are considered,  based on either using a single  or two copies of entangled atom-field state, respectively. It is shown that, by using the first version, the fidelity of the teleported state as well as the amount of quantum Fisher information, that contains in the teleported state, are  much better than using the second version. In general, one may increases the fidelity of teleported information by increasing the mean photon number and decreasing the detuning parameter. The fidelity of teleporting classical information is much better than teleporting quantum information. Moreover, teleportating classical information that initially encoded in an exited states is much better than that encodes in the ground states. However, the teleported Fisher information that initially encoded in a ground state is much larger than those initially  encoded in entangled states.

\end{abstract}

\section{Introduction}

Since the first proposed protocol of quantum teleportation by Bennet $et.$ $al.$ \cite{Bennet}, many suggested protocols  are suggested  in both axes theoretically and experimentally. Effectively, their applications are very large, especially in the context of  quantum communication and  information theory \cite{1,2}. The aim of these protocols is  transferring an unknown quantum state remotely with all security in contrast to the classical information theory \cite{2}. Thanks to "non-cloning theorem" \cite{3} which introduced by Wootters $el.$ $al.$ in 1982, it is impossible to clone or to get an identical copy of an arbitrary quantum state. Recently, Ortigoso \cite{4} denoted that a complete proof in terms of the lack of simple nondisturbing measurements in quantum mechanics, also it was already delivered by Park in 1970 \cite{5}. In the same way, Herbert wrote an interesting paper which used the non cloning theorem to propose a superluminal communication device by using quantum entanglement \cite{6}. This type of quantum correlations is caused by the
non-local hidden variables, named "Entanglement" by
Schrodinger \cite{Schodinger}. Bell theorem \cite{2} ensured that two interacting quantum systems in the past can no longer be regarded as independent systems, one can say that they are entangled. Indeed, the entanglement can be seen as the physical phenomenon which occurs when a quantum state of two or more particles interact in such way the quantum state of each particle cannot be described independently of others.\\

In general, entanglement is a fundamental concept in quantum communication such as quantum cryptography and quantum teleportation \cite{2}. The last one uses two kind of channels, the first is a quantum channel which allows to transfer an unknown state and
the second is a classical channel which helps to complete the communication. However, the quantum channel can be either a partially or a maximally, Bell state \cite{2}, entangled state \cite{allati15,allati2011}. For the partially entangled state, it can be arrived for example from an interaction between 2-levels atom with single electromagnetic field. The quality of the quantum information processing may  be
shown by measuring the so called "fidelity" \cite{8}. Furthermore, it gets an idea about how the final state closes to the initial state, i.e the fidelity measures the similarity between the unknown state and the final state at the end of the protocol.\\

In the quantum transmission, the total physical information encoded in a quantum state can not  always be transmitted, but one  can transmit only the relevant parameter information. Moreover, the relevant parameters that carry the information can be exploited in quantum metrology \cite{9}, quantum estimation theory \cite{10}, quantum information theory  and so on. As it is known, the quantification of these parameters can be evaluated by the so called "Quantum Fisher Information" (QFI). This quantifier may  be considered as the most important measures in quantum estimation theory. In general QFI is defined as the sensibility of a quantum state with respect to changes the relevant parameters encoded in this state, i.e. QFI is a quantifier of the precision parameter estimation. Moreover it can be considered as good resource which detect the entanglement between two particles, in particular it it measures the quantitatively the information flow changed between an open quantum state and its surrounding environment. Consequently, quantum Fisher information has attracted attention of many authors in order to evaluate the relevant parameters. Zheng $et.$ $al.$ \cite{11} investigated the dynamics of QFI for a two-qubit system, where each qubit interacts with its own Markovian environment. Ozaydin \cite{12} quantified the QFI analytically for the W-state in the presence of different noisy channels. Metwally \cite{Metwally2018,Metwally17}, discussed the behavior of Fisher information for accelerated and pulsed systems.  Recently, El Anouz $et.$ $al.$ \cite{13} evaluated QFI for a single atomic-field state which used as a quantum channel in quantum teleportation protocol. They showed that it's possible to estimating the teleported initial state parameters in resonance and non-resonance cases.\\

In this paper, we use two schemes of quantum teleportation. In the first scheme ($\mathcal{F}_{TP}$) we  use a single atomic-field state as a quantum channel to  perform the quantum teleportation protocol, while for second scheme ($\mathcal{S}_{TP}$), two copies of the entangled atomic-field state are used to achieve the teleportation protocol. It is shown that, the atomic-field parameters have the same effect on the behavior of the fidelity of the teleported state as well as on the amount of Fisher information that encoded  in the teleported state for both versions;  $\mathcal{F}_{TP}$, and  $\mathcal{S}_{TP}$.
The paper is organized as follows. In the next section we suggest a physical model based on the interaction between two level atom and a single electromagnetic  mode field. In Sec.$3$ a general definitions and notations of the quantum Fisher information are presented. The effect of the field and the atom’s  parameter is displayed in Sec.$4$. Finally our results are summarized in Sec$.5$.

\section{Suggested model}

In the literature, the famous model that describes the interaction matter-light is Jaynes-Cummings model. The Hamiltonian of the suggested system, a single-two level atom  interacts with a single cavity mode in
the rotating wave approximations is given by,
\begin{equation}\label{H}
H= H_0+ \lambda(a|e\rangle \langle g|+
a^{\dagger}|g\rangle \langle e|),
\end{equation}
where $H_0$ represents the free Hamiltonian of the atom
and the cavity mode, $H_0
=\omega_{0}\sigma_{z}/2+\omega a^{\dagger}a$. The second term represents
the interaction Hamiltonian between the atom and the single field with the coupling constant $\lambda$. The frequencies of the atomic system and the cavity
mode are defined by $\omega_0$ and
$\omega$, respectively. The lower and the ground levels of the atom are  $|e\rangle$ and $|g\rangle$, respectively. Moreover, inside the cavity $a^\dagger$ ($a$) describes the creation (the annihilation) operator which allow to move up (down) a quantum state.\\

Let's suppose that, the states of the atomic and the field
represent  by $\ket{\phi_A(0)}=\ket{e}$ and
$\ket{\phi_F(0)}=\sum_{0}^{\infty} P_{n}\ket{n}$, respectively. The atomic-field system is initially defined by,
\begin{equation}
|\phi_{S}(0)\rangle=\sum_{n=0}^{+\infty} P_{n}|n\rangle\otimes |e\rangle,
\end{equation}
where, $P_{n}=\exp(-\frac{\bar{n}}{2}) \sqrt{\frac{\bar{n}^n}{n!}}$, and
$\bar{n}$ is the average photons number. At  an arbitrary time ($t\neq0$), the whole system is given by:
\begin{equation}\label{nn}
|\phi_{S}(t)\rangle=\mathcal{O}(t)|\phi_{S}(0)\rangle,
\end{equation}
with the time  evolution operator $\mathcal{O}(t)=e^{-iHt}$.
In an explicit form, the final state (\ref{nn}) of atomic-field system takes the form,
\begin{eqnarray}\label{psi(t)}
|\phi_S(t)\rangle&=&\sum_{n=0}^{\infty}P_n\Bigl\{\Bigl(\cos(\tau\sqrt{\delta^2+\hat{n}})-
i\frac{\delta}{\sqrt{\delta^2+ \hat{n}}}
\sin(\tau\sqrt{\delta^2+\hat{n}})|n,e\rangle\Bigr) \nonumber\\
&&-i e^{-i\delta \tau}\frac{
\sqrt{n+1}}{\sqrt{\delta^{2}+(n+1)}}~\sin(\tau\sqrt{\delta^2+\hat{n}})|n+1,g\rangle\Bigr\},
\end{eqnarray}
with $\hat{n}=a^{\dagger}a$ is the mean photon number, $\tau=\lambda t$ and
$\delta=\Delta/2\lambda$ are  the dimensionless time and detuning parameters, respectively .
It is clear that, the state Eq.(\ref{psi(t)}) represents the final state of the
whole system in  $2 \times \infty$ -dimensions space. In order to measure the minimum amount of entanglement which is
generated between the atom and the cavity,
the final state Eq.(\ref{nn}) is projected  in a $2\times 2$ dimensions subspace \cite{13}. In
the 2-dimensions space, the atomic-field state is collapses to be,

\begin{eqnarray}\label{final}
\rho_{S}(t)&=&\alpha_{1}|n,g\rangle\langle
n,g|+\alpha_{2}|n,e\rangle\langle
n,e|+\alpha_{3}|n,e\rangle\langle
n+1,g|+\alpha_{3}^{*}|n+1,g\rangle\langle
n,e|\nonumber\\
&&+ \alpha_{4}|n+1,g\rangle\langle
n+1,g|+\alpha_{5}|n+1,e\rangle\langle n+1,e|,
\end{eqnarray}
where the different coefficients $\alpha_i$ ($i=1,...,5$) are  simply obtained from Eq.(4) (see appendix $A$).

\section{Quantum Fisher Information}

In quantum information theory, quantifying directly the parameters that describe the  quantum state is not always possible.
For that, quantum Fisher information (QFI) gets a way to solve this problem by estimating these parameters \cite{15}. QFI enables to measures the sensitivity of quantum state or the physical parameter encoded in this state. For example, let's consider a two-qubits system which depends on parameter $\xi$, the corresponding quantum Fisher information $\mathcal{F}_\xi$  is defined as \cite{16}
\begin{equation}\label{Fisher}
\mathcal{F}_{\xi}=Tr[\rho_{\xi}\mathcal{L_{\xi}^{\rho}}]=Tr[(\partial_{\xi}\rho_{\xi})\mathcal{L_{\xi}}],
\end{equation}
where $\mathcal{L_{\xi}}$ is the symmetric
logarithmic derivative, which is given by
$\partial_{\xi}\rho_{\xi}=(\mathcal{L_{\xi} \rho_{\xi}}+ \rho_{\xi}\mathcal{L_{\xi}})/2$ with
$\partial_{\xi}=\partial/\partial_{\xi}$. Using the spectrum decomposition of   $\rho_{\xi}$ as  $\rho_{\xi}=\sum_{k=1}^{n}\lambda_{k}|\psi_{k}\rangle \langle \psi_{k}|$ where $\lambda_{k}$ and $|\psi_{k}\rangle $ are the eigenvalues and the eigenvectors respectively, one can rewrite the QFI as \cite{Metwally17}
\begin{equation}\label{QFI}
\mathcal{F}_{\xi}=\sum_{k}
\frac{(\partial_{\xi}\lambda_{k})^{2}}{\lambda_{k}}+4\sum_{k}\lambda_{k} [ \langle  \partial_{\xi} \psi_{k}|\partial_{\xi} \psi_{k}\rangle -|\langle\psi_{k}|\partial_{\xi}\psi_{k}\rangle|^{2} ]-\sum_{k\neq
	l} \frac{8\lambda_{k}
	\lambda_{l}}{\lambda_{l}+\lambda_{k}}|\langle\psi_{k}|\partial_{\xi}\psi_{l}\rangle|^{2},
\end{equation}
where, the first term in Eq .(\ref{QFI}) allows to investigate the so called classical Fisher information. However, the possibility to estimate the parameter $\xi$ encoded in the pure  state $|\psi_k \rangle $ is described by the second term in Eq. (\ref{QFI}). While the possibility to estimate the same parameter $\xi$ of a mixed state is given by  the third term in Eq.(\ref{QFI}). In general, the sensitivity of a mixte state is smaller the sensitivity of a pure state.\\

In quantum teleportation protocol, we teleport the full included information in a given quantum state, where the efficiency scheme is measured by the fidelity. Recently it shown that it is enough to teleport some relevant included parameters in the quantum state, and the teleportation  credibility will be investigated by quantum Fisher with respect to these parameters. Indeed in following section, we shall compare the fidelity and the quantum Fisher of a teleported state using two different teleportation schemes.

\section{Teleportation protocol}

It is well known that, the quantum teleportation is one of the most important achievements over the last decade. However, it's almost the engine of all the tasks in the quantum information processing. In  this section, the generated entangled atomic-field state (\ref{final}) is employed to teleport an unknown state between two legitime users; Alice and Bob who share the state (\ref{final}). Let's assume that, Alice want to send the following unknown state
\begin{equation}\label{unk}
\ket{\psi_{un}}=\cos\theta\ket{00}+ e^{-i\phi/2}\sin\theta\ket{11},
\end{equation}
where $\theta\in[0,\pi]$ and  $ \varphi\in[0,2\pi]$. Next, we propose two suggested schemes of quantum teleportation for achieving the communication between the users. Now Alice's aim is to send the state Eq.(\ref{unk}) to Bob. For this task, Alice may use the first teleportation version, $\mathcal{F}_{TP}$, in which the partners use only  a single copy of atomic-field state (\ref{final}), or they  use the second version $\mathcal{S}_{TP}$, where two copies of the state (\ref{final}) are used.

\subsection{First Teleportation Protocol $\mathcal{F}_{TP}$}

In this suggested protocol, it is assumed that the partners, Alice and Bob share a single copy of the state (\ref{final}). Alice is given unknown  state (\ref{unk}). The users perform the steps of the teleportation protocol  which are  described  in   \cite{13}. At the end of the protocol,  Bob will get the state
\begin{equation}
\rho_{B_1}=\sum_{i j}P_{i j}(\sigma_{i}\otimes\sigma_{j})\rho_{un}(\sigma_{j}\otimes\sigma_{i}),
\end{equation}
where $P_{ij}=Tr[E^{i}\rho_{ac}]Tr[E^{j}\rho_{ac}]$, $
\sum_{ij}P_{ij}=1$ ($ij=0,x,y,z$), and  $\sigma_{m}$ ($m=i, j$) are the Pauli operators, the different projection $E$ are given by
\begin{eqnarray}
E^{0, z}&=&|\psi^{\mp}\rangle\langle\psi^{\mp}|, ~~\quad\quad\quad\quad
 E^{x, y}=|\phi^{\mp}\rangle\langle\phi^{\mp}|, \nonumber\\
 |\psi^{\pm}\rangle&=&\frac{1}{\sqrt{2}}({|01\rangle+|10\rangle}),\quad\quad
 |\phi^{\pm}\rangle=\frac{1}{\sqrt{2}}({|00\rangle+|11\rangle}),
\end{eqnarray}
where
$|\psi^{\pm}\rangle$ and $
|\phi^{\pm}\rangle$ represent the Bell states \cite{2}. Using these different expressions of states, the final Bob's state in  the basis $\{|n,g\rangle, |n,e\rangle, |n+1,g\rangle, |n+1,e\rangle\}$, is given by
\begin{equation}\label{B1}
\rho_{Bob_1}=\begin{pmatrix}
\beta_{1}  & 0 & 0 & \beta_{2} \\
0 & 0 & \beta_{3} & 0 \\
0 & \beta_{3}^{*} & 0 & 0 \\
\beta_{2}^{*} & 0 & 0 & \beta_{4}
\end{pmatrix} ,
\end{equation}
where,
\begin{eqnarray}
\beta_{1}&=&(\alpha_{1}+\alpha_{5})^2 \sin^2(\theta)+ (\alpha_{2}+\alpha_{4})^2\cos^2(\theta) ,\nonumber\\
\beta_{2}&=& (\alpha_{3}+\alpha_{3}^{*})^{2} \cos(\theta)\sin(\theta)e^{i\varphi/2}, \nonumber\\
\beta_{3}&=& (\alpha_{1}+\alpha_{5}) (\alpha_{2}+\alpha_{4}) , \nonumber\\
\beta_{4}&=& (\alpha_{2}+\alpha_{5})^2 \cos^2(\theta)+ (\alpha_{1}+\alpha_{4})^2\sin^2(\theta),
\end{eqnarray}
and  $\alpha_{i}$ ($i=1,..,5$) are given in   appendix ($A$). The measure of how the final state closes to the initial state is given by the fidelity as,
\begin{eqnarray}
\mathcal{F}_{B_1}&=&\Bigl[(\alpha_{1}+\alpha_{5})^2 \sin^2(\theta)+ (\alpha_{2}+\alpha_{4})^2\cos^2(\theta)\Bigr] \cos^2(\theta)+ (\alpha_{3}+\alpha_{3}^{*})^{2} \sin^2(2\theta)/2
\nonumber\\
 &&+\Bigl[(\alpha_{2}+\alpha_{5})^2 \cos^2(\theta)+ (\alpha_{1}+\alpha_{4})^2\sin^2(\theta)\Bigr] \sin^2(\theta).
 \end{eqnarray}

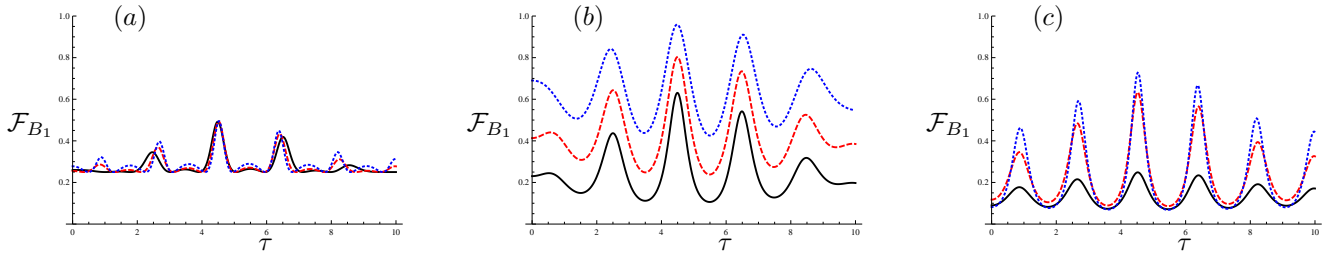
\begin{figure}[h!]
	\begin{center}
	\includegraphics[scale=.5]{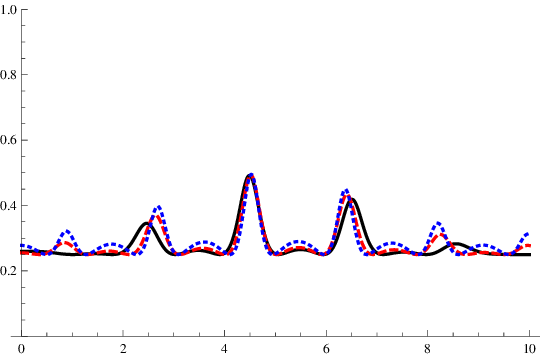}
	\put(-150,40){$\mathcal{F}_{B_1}$}
	\put(-55,-5){$\tau$}
\put(-110,80){$(a)$}~~~~\quad\quad\quad
	\includegraphics[scale=.5]{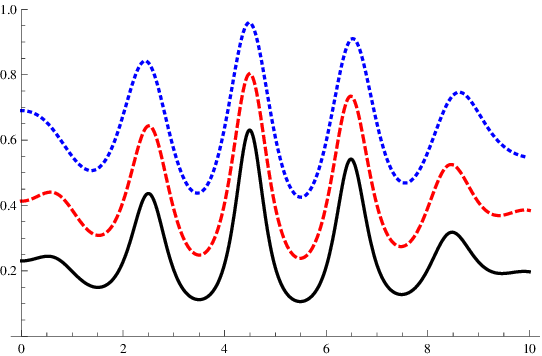}
	\put(-150,40){$\mathcal{F}_{B_1}$}
	\put(-55,-5){$\tau$}
\put(-110,80){$(b)$}~~~~\quad \quad\quad
	\includegraphics[scale=.5]{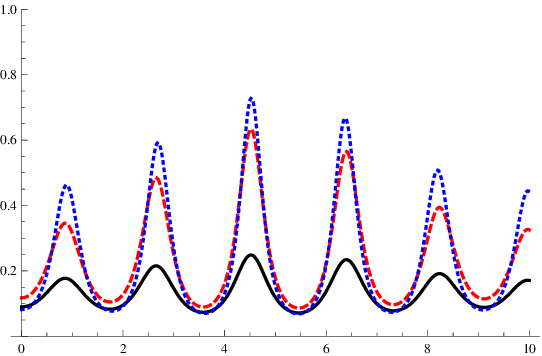}
	\put(-150,40){$\mathcal{F}_{B_1}$}
	\put(-55,-5){$\tau$}
\put(-110,80){$(c)$}
		\caption{ The fidelity of the teleported state (\ref{unk}) by using a single copy of (\ref{final}) in the resonance case $\delta=0$,$n=2$, and $\phi=0$, where 		(a) $\theta=\pi/4$, (b) $\theta=\pi/2$ and (c) $\theta=0$. The solid, dash and dot for  $\bar{n}=2, 4, 6$, respectively.}
	\end{center}
\end{figure}

Fig.(1) displays the behavior of the fidelity $\mathcal{F}_{B_1}$ at  the resonance case ($\delta=0$), where the teleported state
is initially prepared in the state (\ref{unk}) and different values of the $\bar{n}$  are considered. The general behavior
shows that, the fidelity $\mathcal{F}_{B_1}$, oscillates between its upper and lower bounds. The maximum/minumm values
depend on the value of the mean photon numbers $\bar{n}$, inside the cavity, where as $\bar{n}$ increases, the maximum
bounds increase. The amplitude of oscillation depends on the type of the teleported information. However,
as it is shown in Fig.(1a), where the initial state prepared in the state $|\psi_{un} \rangle =\frac{1}{\sqrt{2}} ( |00 \rangle + |11 \rangle)$, the amplitude of these  oscillations is small and the minimum value of the fidelity is slightly changed at different values of the mean photon numbers. Moreover, the fidelity survives during the interaction time and never vanishes. Fig.(1b),
displays the behavior of the fidelity when the initial teleported state contains only classical information,
namely $|\psi_{un}\rangle =\ket{11}$. In this case, the amplitudes of these oscillations are  large and the effect of the mean photon
number is clearly displayed. The Fidelity of the teleported state, may vanish periodically and fast as the
mean photon numbers increases. The fidelity of teleported unknown state by using $\mathcal{F}_{TP}$, when the initial teleported state is prepared in the ground state is displayed in Fig.(1c). The behavior of $\mathcal{F}_{B_1}$ is similar to that predicted in Fig.(1b), but the upper bounds are much smaller. The mean photon number has the same effect on the behavior of the fidelity, where the fidelity increases as $\bar n$ increases.\\

From Fig.(1), by using the $\mathcal{F}_{TP}$, the possibility of teleporting  unknown state depends on the initial state settings, where the maximum fidelity is predicted if the initial information is encoded in an excited state. The minimum bounds of teleporting quantum information are much better than teleporting classical information. Therefore, one can conclude that, the maximum fidelity is achieved at large values of the  mean photon number and encoding the unknown information in an excite state.\\

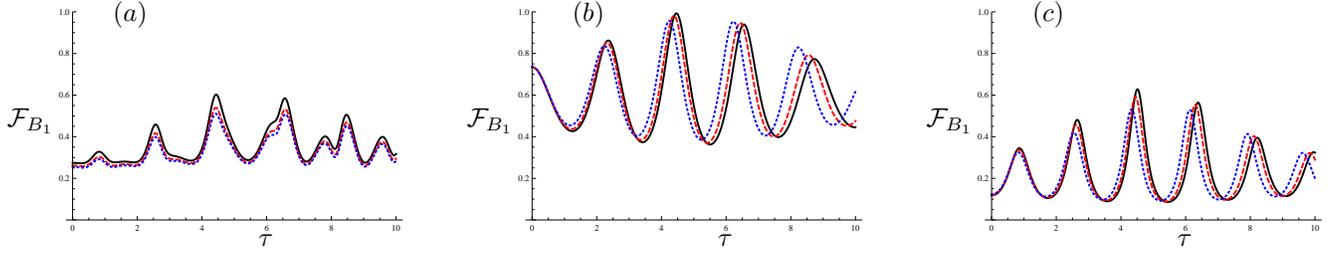
\begin{figure}
	\begin{center}
\includegraphics[scale=.5]{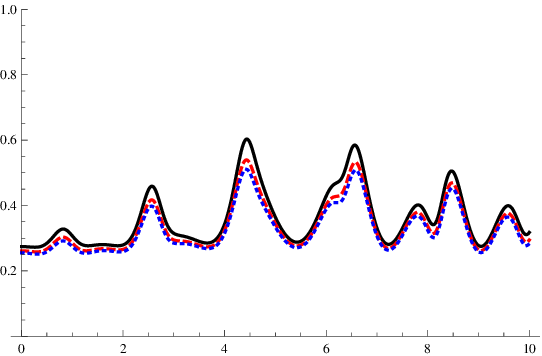}
\put(-150,40){$\mathcal{F}_{B_1}$}
	\put(-55,-5){$\tau$}
\put(-110,80){$(a)$}~~~~\quad\quad\quad
\includegraphics[scale=.5]{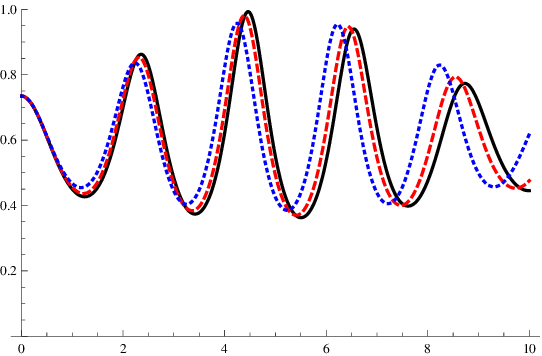}
\put(-150,40){$\mathcal{F}_{B_1}$}
	\put(-55,-5){$\tau$}
\put(-110,80){$(b)$}~~~~\quad \quad\quad
\includegraphics[scale=.5]{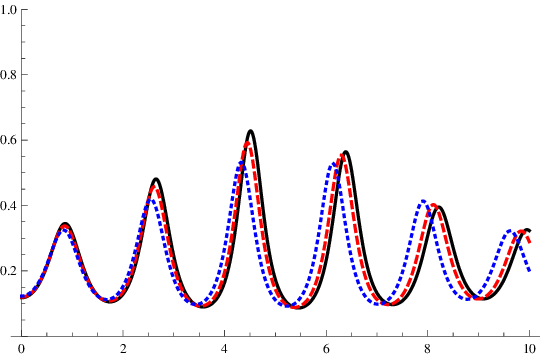}
\put(-150,40){$\mathcal{F}_{B_1}$}
	\put(-55,-5){$\tau$}
\put(-110,80){$(c)$}
		\caption{ The same as Fig.(1) but for non resonance case, where the solid, dash and dot curves for
		$\delta=0.1, 0.3, 0.5$ respectively, n=2 and $\bar{n}=4$.}
	\end{center}
\end{figure}
In Fig.(2), we investigate the behavior of the fidelity of the teleported information at non-resonance
case, namely $\delta \neq 0$. The displayed behavior is similar to that depicted in Fig.(1). The different values of
$\delta$ has  a remarkable  effect on the maximum/minimum  bounds of the fidelity of teleported state, where these  bounds
are  much better at non-resonance case.  Small values of $\delta$ improve the minimum and the maximum values of the fidelity. Also, the  teleported information that encodes on the unknown state plays an important role on its teleported fidelity. However, teleporting classical information prepared initially in an excited state is much better than those prepared in  a ground  state. Although, the fidelity of teleporting quantum information is improved in the non-resonance case, but it  is smaller than those for  classical information, which are initially encoded in the excited state.\\

From Fig.(1) and (2), we conclude that small values of the detuning parameter improves the minimum bounds of the teleported fidelity. Different values of the detuning parameter have slightly effect on the upper bounds of the  fidelity of the teleported state.\\

 \begin{figure}
 	\begin{center}
 		\includegraphics[scale=.5]{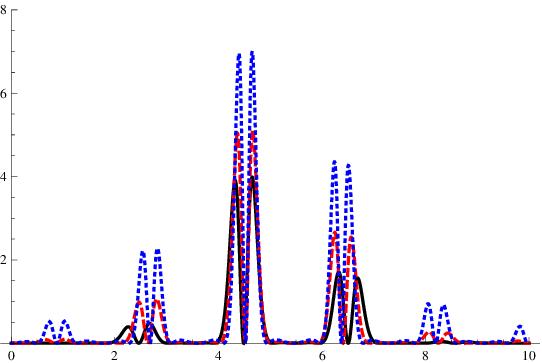}
 	\put(-158,40){$\mathcal{F}_{I_1}(\theta)$}
	\put(-55,-5){$\tau$}
\put(-110,80){$(a)$}~~~~\quad\quad\quad
 		\includegraphics[scale=.5]{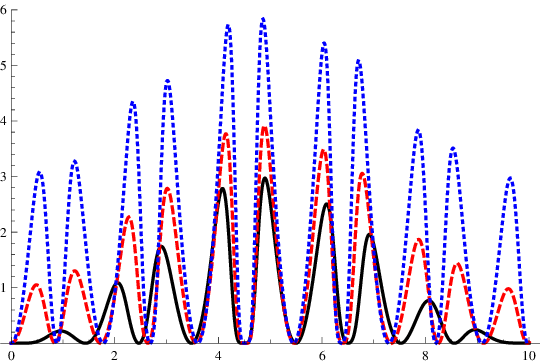}
 		\put(-158,40){$\mathcal{F}_{I_1}(\theta)$}
	\put(-55,-5){$\tau$}
\put(-110,80){$(b)$}~~~~\quad \quad\quad
 		\includegraphics[scale=.5]{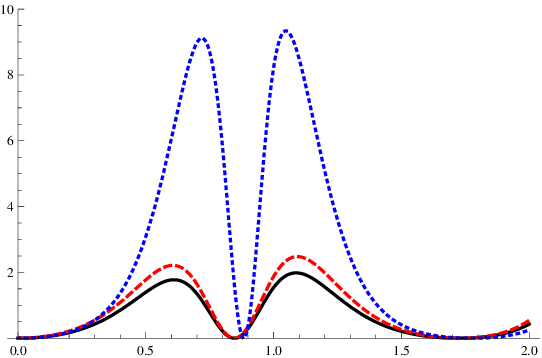}
 		\put(-158,40){$\mathcal{F}_{I_1}(\theta)$}
	\put(-55,-5){$\tau$}
\put(-110,80){$(c)$}
 		\caption{The  Fisher information  with respect to the parameter $\theta$, $\mathcal{F}_{I_1}(\theta)$ of the state (\ref{B1}) at the resonance case $\delta=0$,$n = 2$, where (a) $\theta=\pi/4$, (b) $\theta=\pi/2$ and (c) $\theta=0$. The solid, dash and dot for $\bar{n}=2,4,6$, respectively.}
 	\end{center}
 \end{figure}

 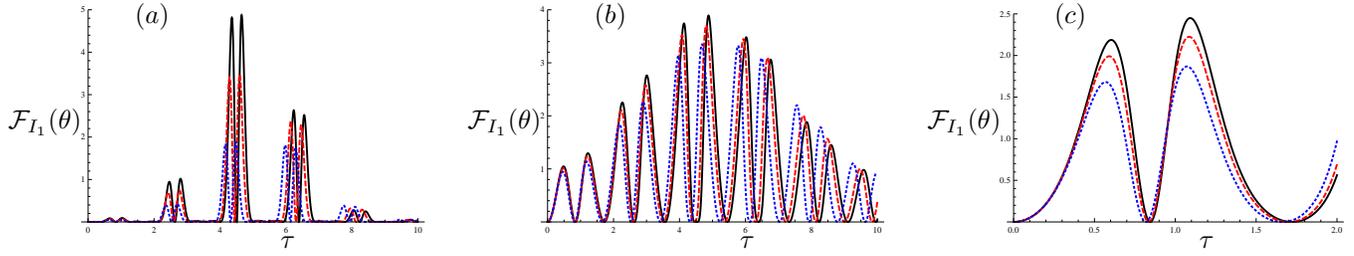
\begin{figure}
	\begin{center}
			\includegraphics[scale=.5]{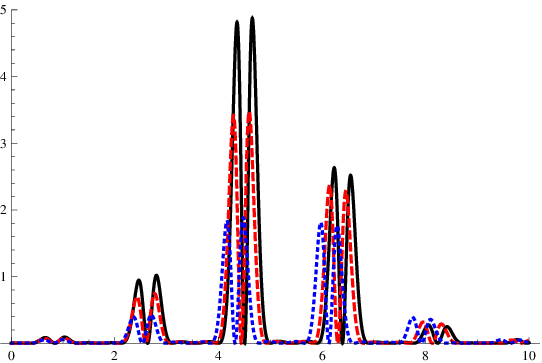}
		\put(-158,40){$\mathcal{F}_{I_1}(\theta)$}
	\put(-55,-5){$\tau$}
\put(-110,80){$(a)$}~~~~\quad\quad\quad
		\includegraphics[scale=.5]{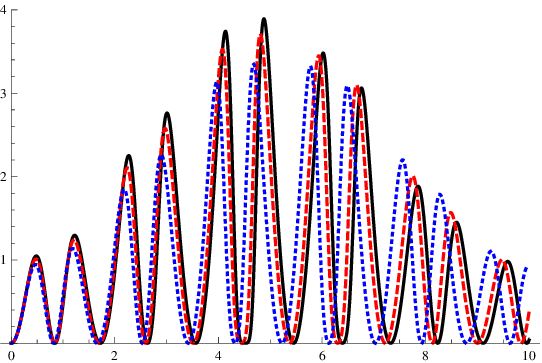}
		\put(-158,40){$\mathcal{F}_{I_1}(\theta)$}
	\put(-55,-5){$\tau$}
\put(-110,80){$(b)$}~~~~\quad \quad\quad
		\includegraphics[scale=.5]{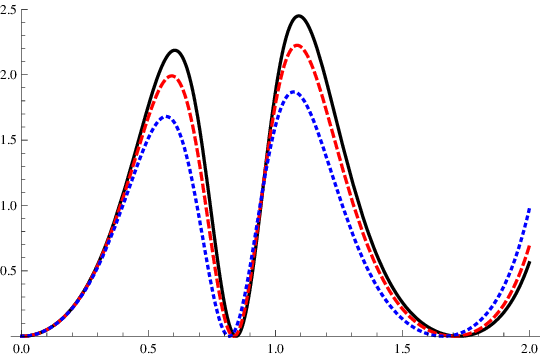}
			\put(-158,40){$\mathcal{F}_{I_1}(\theta)$}
	\put(-55,-5){$\tau$}
\put(-110,80){$(c)$}
		\caption{The same as Fig.(3) but for non resonance case, where the solid, dash and dot curves for
			$\delta=0.1, 0.3, 0.5$ respectively, n=2 and $\bar{n}=4$.}
	\end{center}
\end{figure}

The amount of Fisher information $\mathcal{F}_{I_1}(\theta)$, with respect to the parameter $\theta$, that contained on the teported state is displayed in Figs.(3) and (4). Similarly, the   behavior of  $\mathcal{F}_{I_1}(\theta)$ depends on the type of the teleported information.
However, as it is shown in Fig.(3a), the teleported $\mathcal{F}_{I_1}(\theta)$ that encoded in the state  $|\psi_{un}(\theta=\pi/4)\rangle $ collapses, revivals and death periodically. The behavior of teleported  quantum Fisher information  that encoded in  $|\psi_{un}(\theta=0) \rangle $ is displayed in Fig.(3b). In this case, the behavior of Fisher information  is different from that depicted in Fig.(3a), where the $\mathcal{F}_{I_1}(\theta)$ oscillates between the upper bounds and temporary vanishes. Moreover, the upper bounds of Fisher information  increase as the mean photon number increases.  In Fig.(3c), we plot the quantum Fisher information $\mathcal{F}_{I_1}(\theta)$, when the initial information is encoded in a ground state. The general behavior is similar to that displayed in Fig.(3b), but the maximum bounds are very large compared with those displayed in Figs.(3a),(3b). Moreover, it increases gradually to reach its maximum values and vanishes temporary at large time compared with that displayed in Figs.(3a),(3b).\\

The effect of the detuning parameter on the teleported Fisher information is displayed in Fig.(4). A
similar behavior is shown as that depicted in Fig.(3). However, as one increases the detuning parameter
the amount of the telported quantum Fisher information  decreases and reaches its minimum values fast. Similarly, the maximum
amount of the teleported $\mathcal{F}_{I_1}(\theta)$ depends on the initial type of the encoded  information. However, as it is displayed in Fig.(3a) the minimum amount of the teleported quantum Fisher information is depicted  if the initial information is encoded with $\theta=\pi/4$, where $\mathcal{F}_{I_1}(\theta)$ disappears for long time. It is clear that, the larger values of the detuning parameter, increases the fidelity of teleported Fisher information that encoded on the ground state. At the non-resonance case, the upper bounds of  $\mathcal{F}_{I_1}(\theta)$ increases at small value of the detuning parameter. Moreover, the maximum bounds are very large than those displayed in Figs.(4b) and (4c), where the initial information is encoded in the states $\ket{\psi_{un}(\pi/4)}$ and $\ket{\psi_{un}(\pi/2)}$, respectively.
\begin{figure}
	\begin{center}
	\includegraphics[scale=.5]{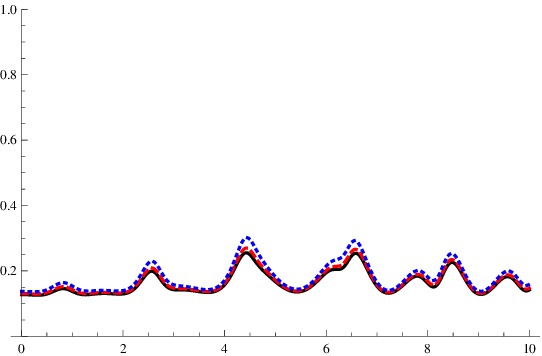}
	\put(-150,40){$\mathcal{F}_{B_2}$}
	\put(-55,-5){$\tau$}
\put(-110,80){$(a)$}~~~~\quad\quad\quad
	\includegraphics[scale=.5]{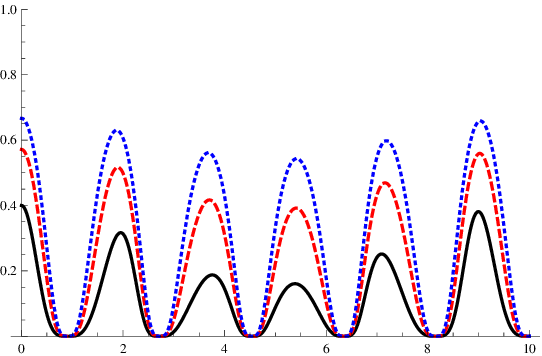}
	\put(-150,40){$\mathcal{F}_{B_2}$}
	\put(-55,-5){$\tau$}
\put(-110,80){$(b)$}~~~~\quad\quad\quad
	\includegraphics[scale=.5]{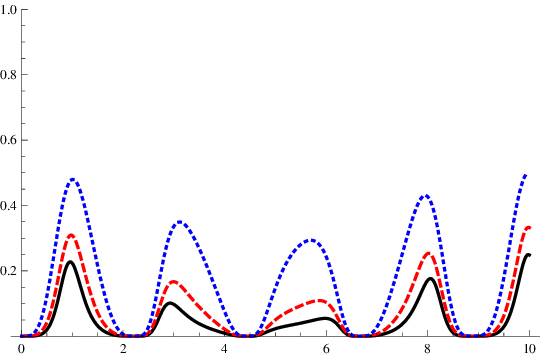}
	\put(-150,40){$\mathcal{F}_{B_2}$}
	\put(-55,-5){$\tau$}
\put(-110,80){$(c)$}~~~~\quad\quad\quad
		\caption{The same as Fig.(1), but the second teleportation protocol, $\mathcal{S}_{TP}$ is used.}
	\end{center}
\end{figure}
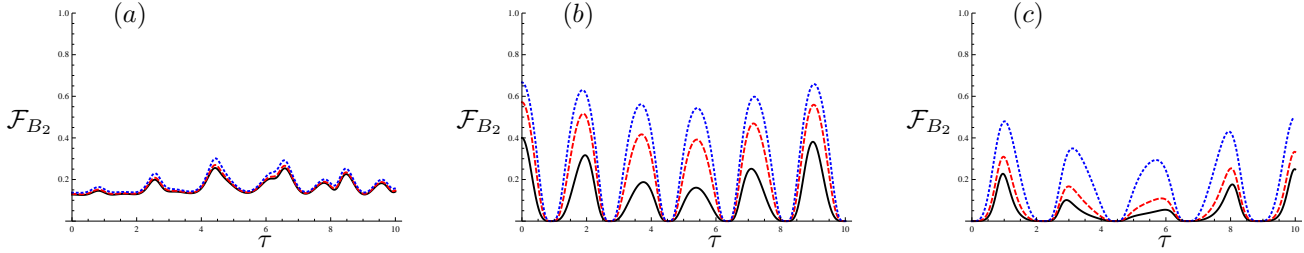

\subsection{Second Teleportation Protocol ($\mathcal{S}_{TP}$)}

In the  second scheme of quantum teleportation $\mathcal{S}_{TP}$, the users share two copies of the state Eq.(\ref{final}). After performing the required steps, Bob will get the state

\begin{equation}\label{FB2}
\rho_{Bob_2}=\begin{pmatrix}
\gamma_1 & 0 & 0 & \gamma_4\\
0 & \gamma_2 & 0 & 0 \\
0 & 0 &\gamma_2 & 0 \\
\gamma_{4}^{*} & 0 & 0 & \gamma_3
\end{pmatrix} ,
\end{equation}
where,
\begin{eqnarray}
\gamma_1 &=& \cos^2(\theta) \alpha_{1}^2 +\sin^2(\theta)\alpha_{4}^2,\nonumber\\
\gamma_2 &=& \cos^2(\theta)  \alpha_{1} \alpha_{2} +\sin^2(\theta)\alpha_{4}\alpha_{5},\nonumber\\
\gamma_3 &=& \cos^2(\theta)  \alpha_{2}^2 +\sin^2(\theta)\alpha_{5}^2,\nonumber\\
\gamma_4 &=& \cos(\theta) \sin(\theta) e^{-i\varphi/2} |\alpha_{3}|^2.
\end{eqnarray}
The fidelity of Bob's state (\ref{FB2}) is given as
\begin{equation}
\mathcal{F}_{B_2}=\cos^2(\theta) [\cos^2(\theta) \alpha_{1}^2 +\sin^2(\theta)\alpha_{4}^2]+\sin^2(\theta) [\cos^2(\theta)  \alpha_{2}^2 +\sin^2(\theta)\alpha_{5}^2]+ 4\cos^2(\theta) \sin^2(\theta) (\alpha_{3}^2+\alpha_{3}^*{^2}).
\end{equation}
\begin{figure}[t!]
	\begin{center}
		\includegraphics[scale=.5]{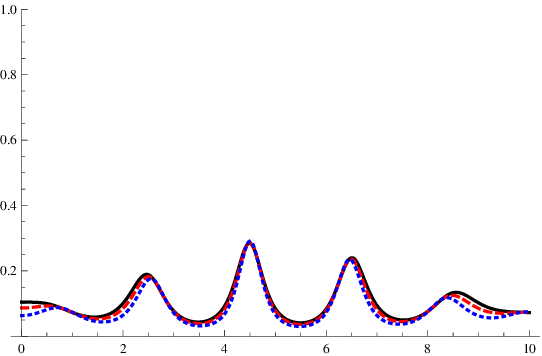}
		\put(-150,40){$\mathcal{F}_{B_2}$}
	\put(-55,-5){$\tau$}
\put(-110,80){$(a)$}~~~~\quad\quad\quad
		\includegraphics[scale=.5]{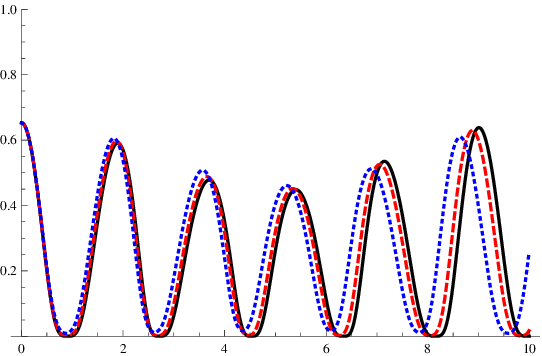}
		\put(-150,40){$\mathcal{F}_{B_2}$}
	\put(-55,-5){$\tau$}
\put(-110,80){$(b)$}~~~~\quad\quad\quad
		\includegraphics[scale=.5]{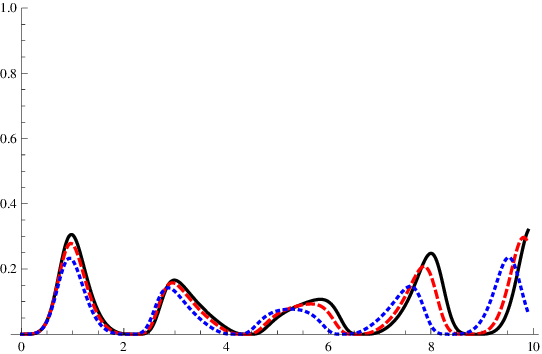}
		\put(-150,40){$\mathcal{F}_{B_2}$}
	\put(-55,-5){$\tau$}
\put(-110,80){$(c)$}~~~~\quad\quad\quad
		\caption{The same as Fig.(5), but for the non-resonance, $\bar{n}=4$.}
	\end{center}
\end{figure}

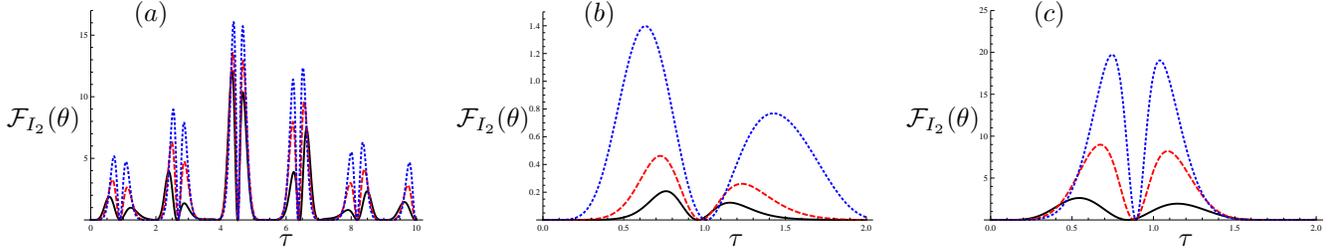
\begin{figure}
	\begin{center}
		\includegraphics[scale=.5]{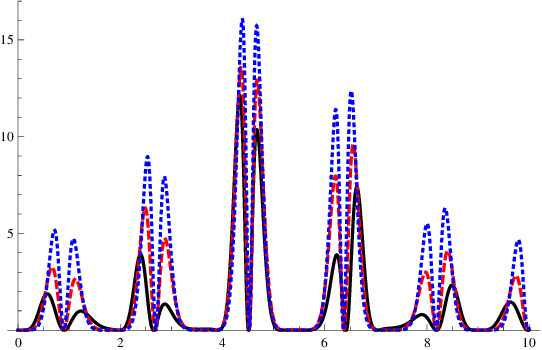}
		\put(-158,40){$\mathcal{F}_{I_2}(\theta)$}
	\put(-55,-5){$\tau$}
\put(-110,80){$(a)$}~~~~\quad\quad\quad
\includegraphics[scale=.5]{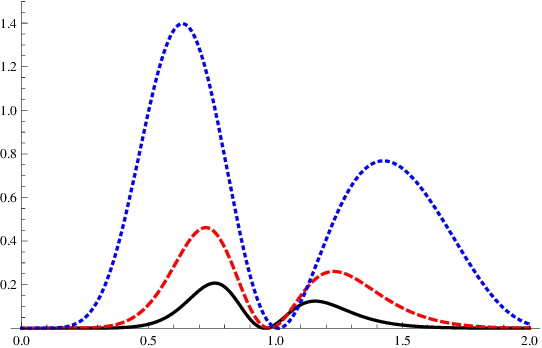}
		\put(-158,40){$\mathcal{F}_{I_2}(\theta)$}
	\put(-55,-5){$\tau$}
\put(-110,80){$(b)$}~~~~\quad\quad\quad
		\includegraphics[scale=.5]{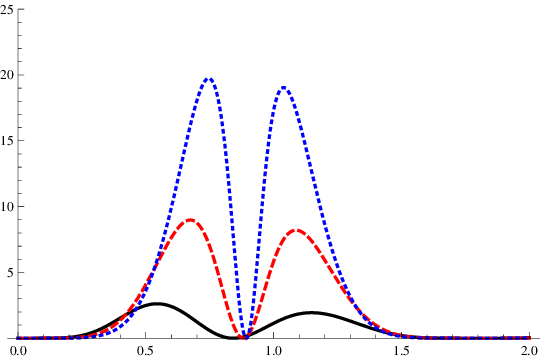}
		\put(-158,40){$\mathcal{F}_{I_2}(\theta)$}
	\put(-55,-5){$\tau$}
\put(-110,80){$(c)$}~~~~\quad\quad\quad
				\caption{The  Fisher information  with respect to the parameter $\theta$, $\mathcal{F}_{I_2}(\theta)$ using the second protocol in resonance case $\delta=0$, $n = 2$, where (a) $\theta=\pi/4$, (b) $\theta=\pi/2$ and (c) $\theta=0$. The solid, dash and dot for $\bar{n}=2,4,6$, respectively.}
	\end{center}
\end{figure}
In Fig.(5), we display the behavior of the fidelity of the teleported state by using the second suggested teleportation protocol, namely the partner share two copies of the atomic-field state (\ref{final}).
The general behavior of $\mathcal{F}_{B_2}$ is similar to that displayed in Fig.(1) for $\mathcal{F}_{B_1}$. The behavior shows that $\mathcal{F}_{B_2}$  flocculates between its maximum and minimum values regularly. The
mean photon number has the same effect that displayed in Fig.(1), namely, the fidelity increases as the mean photon number increases. Also, the initial state settings has a clear effect, where for teleporting classical information the fidelity is much larger than teleporting quantum information. These results are displayed clearly by comparing Fig.(4a) with Figs.(4b) and (4c). Moreover, the minimum values that predicted for teleporting quantum information are much larger than that displayed for teleporting classical information.\\

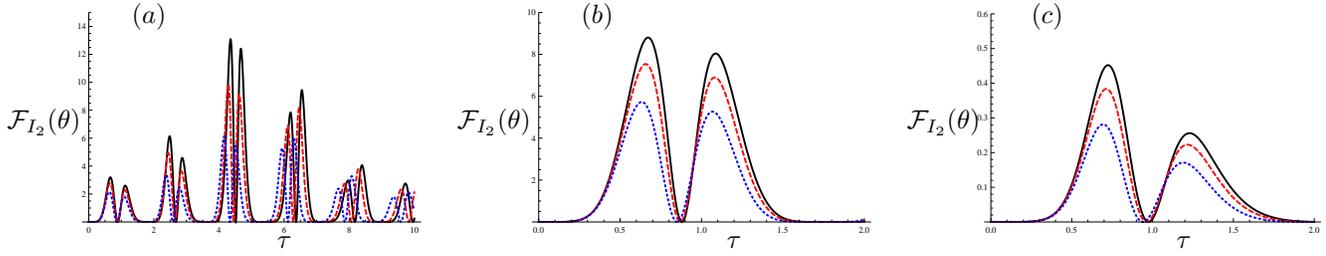
\begin{figure}
	\begin{center}
		\includegraphics[scale=.5]{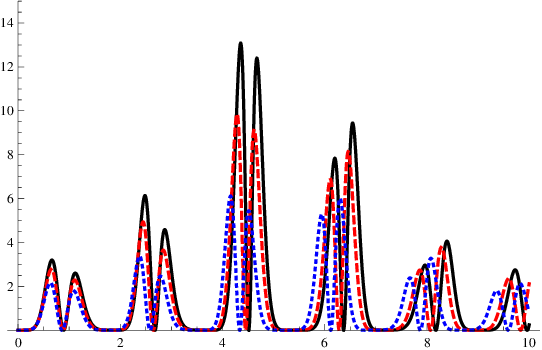}
		\put(-157,40){$\mathcal{F}_{I_2}(\theta)$}
	\put(-55,-5){$\tau$}
\put(-110,80){$(a)$}~~~~\quad\quad\quad
\includegraphics[scale=.5]{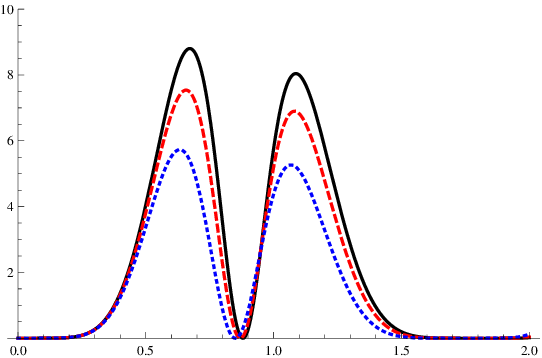}
		\put(-157,40){$\mathcal{F}_{I_2}(\theta)$}
	\put(-55,-5){$\tau$}
\put(-110,80){$(b)$}~~~~\quad\quad\quad
		\includegraphics[scale=.5]{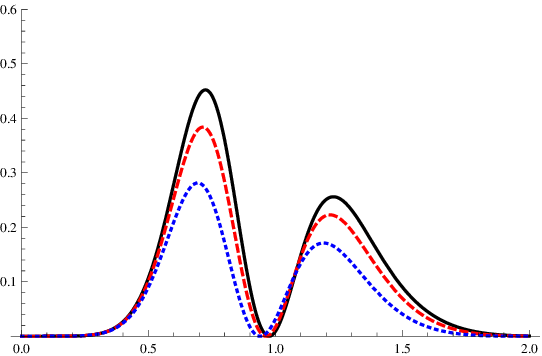}
		\put(-157,40){$\mathcal{F}_{I_2}(\theta)$}
	\put(-55,-5){$\tau$}
\put(-110,80){$(c)$}~~~~\quad\quad\quad
				\caption{ The same as Fig.(7) but for the non-resonance case.}
	\end{center}
\end{figure}

Similarly in Fig.(6), we investigate the behavior of the fidelity $\mathcal{F}_{B_2}$ of the teleported information at non-resonance
case. It is clear that, the upper bounds are larger than those displayed in Fig.(5) (see at $\bar{n}=4$). However, the initial unknown state  settings  has the same effect as that displayed for $\mathcal{F}_{TP}$, where the fidelity of  teleporting quantum information is much smaller than teleporting  classical information. Larger values of the detuning parameter has a slightly effect on the upper bounds of the fidelity, but  increases the exchange between its maximum and minimum values.\\

The quantum Fisher information  $\mathcal{F}_{I_2}(\theta)$ of the teleported state  by using the second version $\mathcal{S}_{TP}$  is displayed in Figs.(7) and (8) for the resonance and non-resonance cases, respectively. The effect of the mean photon number and the detunig parameter is similar to that displayed in Figs.(3) and (4) for $\mathcal{F}_{I_1}(\theta)$, where the partners use the first teleportation version $\mathcal{S}_{TP}$.
The maximum bounds are predicted for the quantum Fisher information that initially encoded in ground state.

\begin{figure}[h!]
	\begin{center}
		\includegraphics[scale=.5]{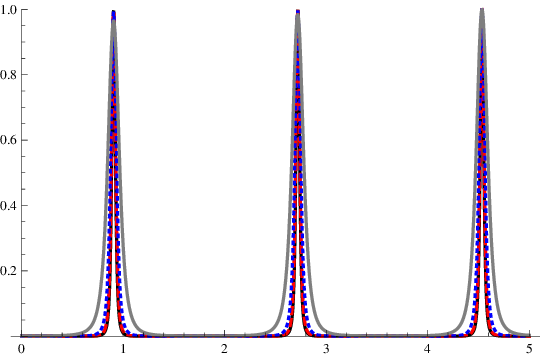}
		\put(-150,40){$\mathcal{F}_{B_1}$}
		\put(-55,-5){$\tau$}
		\put(-120,80){$(a)$}~~~~\quad\quad\quad
		\includegraphics[scale=.5]{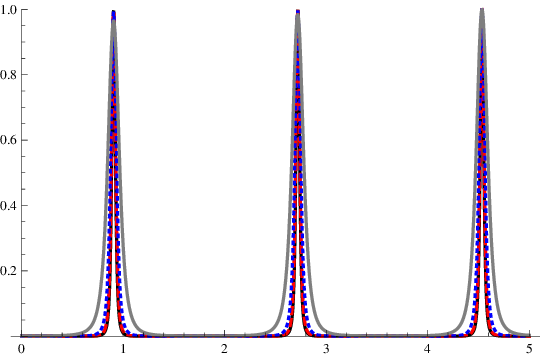}
		\put(-150,40){$\mathcal{F}_{B_1}$}
		\put(-55,-5){$\tau$}
		\put(-120,80){$(b)$}~~~~\quad \quad\quad
		\includegraphics[scale=.5]{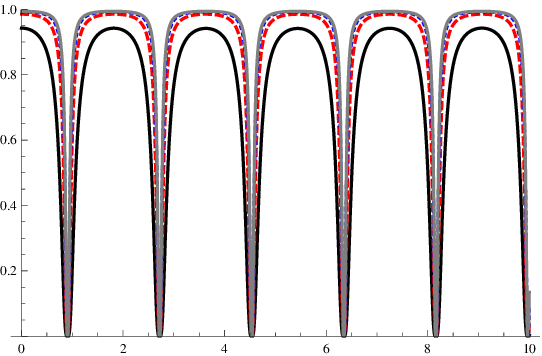}
		\put(-150,40){$\mathcal{F}_{B_1}$}
		\put(-55,-5){$\tau$}
		\put(-150,80){$(c)$}
		\caption{The fidelity of the teleported state, where the $\mathcal{F}_{TP}$ is considered in (a), (b) and $\mathcal{S}_{TP}$ is considered in (c)  in the resonance case $\delta=0$,$n=2$, and $\phi=0$, where 		(a), (c) $\theta=\pi/2$ and (b) $\theta=0$. The Cray, Blue, Red and Black for  $\bar{n}=1000, 800, 400, 100$, respectively.}
	\end{center}
\end{figure}

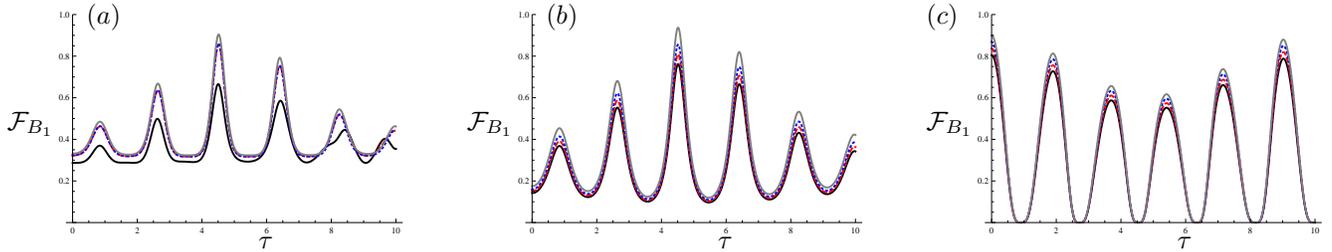
\begin{figure}[h!]
	\begin{center}
		\includegraphics[scale=.5]{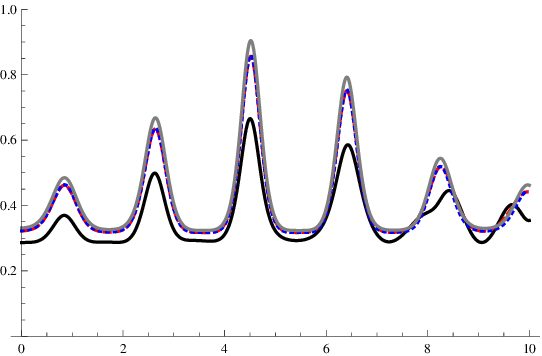}
		\put(-150,40){$\mathcal{F}_{B_1}$}
		\put(-55,-5){$\tau$}
		\put(-120,80){$(a)$}~~~~\quad\quad\quad
		\includegraphics[scale=.5]{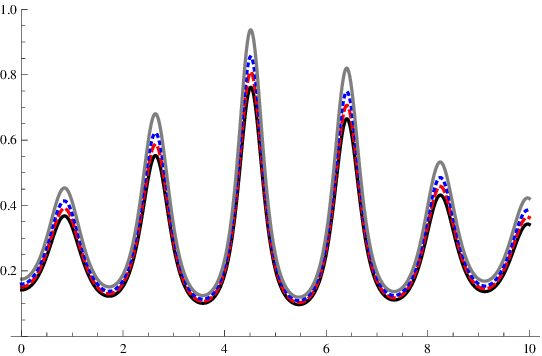}
		\put(-150,40){$\mathcal{F}_{B_1}$}
		\put(-55,-5){$\tau$}
		\put(-120,80){$(b)$}~~~~\quad \quad\quad
		\includegraphics[scale=.5]{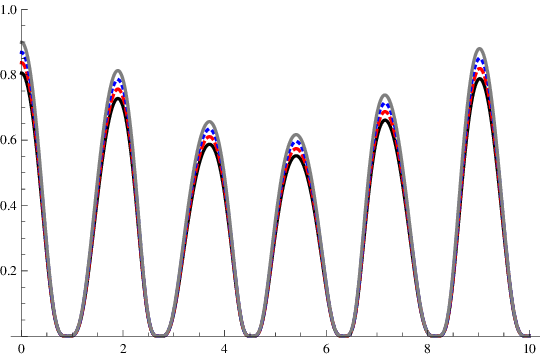}
		\put(-150,40){$\mathcal{F}_{B_1}$}
		\put(-55,-5){$\tau$}
		\put(-150,80){$(c)$}
		\caption{The fidelity of the teleported state, where the $\mathcal{F}_{TP}$ is considered in (a), (b) and $\mathcal{S}_{TP}$ is considered in (c), where $n=2$, $\bar{n}=4$, and $\phi=0$, where (a), (c) refer to $\theta=\pi/4$ and (b) $\theta=0$. The Cray, Blue, Red and Black for  $\delta=0.001, 0.005, 0.02, 0.05$, respectively.}
	\end{center}
\end{figure}

	Moreover in order to maximize the fidelity to one, we consider large number of $\bar{n}$ and small values of $\delta$. Indeed in Fig. (9) and (10) we display the evolution of fidelity as a function of $\tau$. It is clear that in both techniques, namely $\mathcal{F}_{TP}$ and $\mathcal{S}_{TP}$, the fidelity reaches the maximum bounds for large values of the mean number photon and small values of detuning parameters ( $\bar{n}=1000$, $\delta=0.001$)
.Based on this, the control of these parameters gives rise to a  threshold  values of $\bar{n}$ and $\delta$, namely  $\bar{n}=1000$ and  $\delta=0.001$. Thanks to these critical parameters, the fidelity in the first technique $\mathcal{F}_{TP}$  reaches the maximum bound and never exceeds one while in the second  technique $\mathcal{S}_{TP}$ the fidelity approaches to one. However in both  techniques the choice of $\bar{n}\ge 1000$ and $\delta\le 0.001$ gives always the same results where the fidelity still constant or approaches to its maximum bound.

As a result the main advantages of this work is to propose a two schemes of quantum teleportation process, using a single copy  and two copy  of $\ref{psi(t)}$, respectively. The control of the mean number photon and detuning parameters allows to determine a threshold, where the fidelity is maximized to one. Moreover in the proposed models the measures of the sensitivity of the teleported state via quantum Fisher information is investigated in order to improve the credibility of transmitting the information encoded in the unknown quantum state. Using a single copy of the entangled atom-field state as quantum channel ($\mathcal{F}_{TP}$) gives rise to large bound of fidelity as well as high sensitivity, with respect to the threshold value mentioned in Figures 9 and 10. These results may be useful analytically and experimentally in the context of quantum information theory, quantum sensing, quantum metrology and many others fields.

\section{Conclusion}

In this contribution, we investigate the possibility of using the minimum amount of entangled state, that
is generated between an atom and a cavity mode,
to teleport a quantum and classical information encoded in a two-qubit state. In this context, we used
two techniques of a teleportation protocol, in the first one $\mathcal{F}_{TP}$,  the partners share only one copy of the generated entangled atom-field state. However, for the second suggested protocol, $\mathcal{S}_{TP}$, the partners use two copies of the entangled atom-field state. We investigate the effect of the atomic and filed parameters on the fidelity of the
teleported states as well as the amount of the teleported Fisher information. It is shown that,  for both versions $\mathcal{F}_{TP}$ and $\mathcal{S}_{TP}$, the behavior of the teleported information is similar.  However, for  $\mathcal{F}_{TP}$ the teleported state, which contains quantum or classical information  never vanishes, but for the second version $\mathcal{S}_{TP}$, the teleported classical information vanishes periodically. Moreover, the upper bounds of the fidelity as well as the quantum Fisher information that predicted by $\mathcal{F}_{TP}$ is much better than that displayed by the second teleportation version $\mathcal{F}_{TP}$. The  interaction parameters; the mean photon number and detuning parameters have the same effect on both versions.\\

Our results show that the fidelity increases as the mean photon number increases
or the detuning parameter decreases. By these choices, the generated entanglement between the atom and the cavity increases and consequently its efficiency to be used as  quantum channel to perform the telportaion protocols increases. However, the atomic and the field’s parameters have the same effect on the amount of the teleported Fisher
information, namely, the maximum bounds are displayed at large values of the mean photon number
and small values of the detuning parameter. However, the Fisher information that displayed for a state
initially encodes classical information oscillates fast and temporary vanishes to rebirth suddenly. The
freezing phenomena of the Fisher information is displayed if the initial teleported state encodes quantum
information. The initial state settings has a noticeable effect, where the upper
bounds of the fidelity that predicted for a system encodes quantum information are smaller than those
encodes classical information. On the other hand, the classical information may be initially encoded in
an exited or ground states. In this context, the fidelity of the teleported information that encoded in an
excited state is much larger than that encoded initially on the ground state.\\

From the behaviors of the fidelities and the quantum Fisher information, one may conclude that in the resonance and non-resonance  case, the possibility of teleportating quantum/ classical information by using a single copy of a two-qubit state, namely using the first version ($\mathcal{F}_{TP}$) of teleportation, is much better than using the second version ($\mathcal{S}_{TP}$), where two copies of the atomic-field states are used.  However, by using ($\mathcal{F}_{TP}$) version, the fidelity never vanishes and the maximum bounds are always larger than those are  displayed by using the second teleportation version. The detuning and the mean photon numbers have the same effect on the fidelity of the teleported state by using the both versions of the teleportation protocols. In general, one may increases the fidelity of the teleported state by increasing the mean photon number or increasing the detuning parameter. \\

\section*{Appendix A. }

The coefficients $\alpha_{i}$ ($i=0,...,5$) are obtained from Eq.(\ref{psi(t)}) as
\begin{eqnarray}\label{alpha_i}
\alpha_{1}&=&
\frac{P_{n-1}^{2}}{\mathcal{N}}\frac{n}{\delta^2+n}\sin^{2}(\tau\sqrt{\delta^2+n}),
\nonumber\\
\alpha_{2}&=&
\frac{P_{n}^{2}}{\mathcal{N}}\Big[\cos^{2}(\tau\sqrt{\delta^2+(n+1)})+
\frac{\delta^{2}}{\delta^2+(n+1)}\sin^{2}(\tau\sqrt{\delta^2+(n+1)})\Big],
\nonumber\\
\alpha_{3}&=&
i\frac{P_{n}^{2}\sqrt{n+1}e^{-i\delta\tau}}{\mathcal{N}
	\sqrt{\delta^2+(n+1)}}\sin(\tau\sqrt{\delta^2+(n+1)})
\Bigl[\cos(\tau\sqrt{\delta^2+(n+1)}) \nonumber\\
&&-i\frac{\delta \sin(\tau\sqrt{\delta^2+(n+1)})}{\sqrt{\delta^2+(n+1)}}\Bigr],\nonumber\\
\alpha_{4}&=& \frac{P_{n}^{2}}{\mathcal{N}}\frac{(n+1)}{\delta^2+(n+1)}\sin^{2}(\tau\sqrt{\delta^2+(n+1)}),\nonumber\\
\alpha_{5}&=&
\frac{P_{n+1}^{2}}{\mathcal{N}}\Bigl[\cos^{2}(\tau\sqrt{\delta^2+(n+1)})+
\frac{\delta^2}{\delta^2+(n+1)}\sin^{2}(\tau\sqrt{\delta^2+(n+1)})\Bigr],
\end{eqnarray}
where $\delta$ is the dimentionless detuning and  the normalization of the state  $\mathcal{N}$  is given by,
\begin{eqnarray}
\mathcal{N}&=&P_{n}^{2}+ \Bigl[\frac{n}{\delta^2+n}\sin^{2}(\tau\sqrt{\delta^2+n})\Bigr]P_{n-1}^2\nonumber\\
 &&+\Bigl[\cos^{2}(\tau\sqrt{\delta^2+(n+1)})+\frac{\delta^2}{\delta^2+(n+1)}\sin^{2}(\tau\sqrt{\delta^2+(n+1)})\Bigr]P_{n+1}^{2}.
 \end{eqnarray}

\end{document}